 \documentclass[final,3p,times,dvipdfmx]{elsarticle}

\usepackage{amssymb}
\usepackage{amsmath} 
\usepackage{graphicx}
\usepackage{color}
\usepackage{comment}

\journal{Physica A}

\begin{document}

\begin{frontmatter}



\title{Positive and negative effects of social impact on evolutionary vaccination game in networks}


\author[label1]{Genki Ichinose}
\author[label2]{Takehiro Kurisaku}

\address[label1]{Department of Mathematical and Systems Engineering, Shizuoka University, Hamamatsu, 432-8561, Japan}
\address[label2]{Anan National College of Technology 265 Aoki Minobayashi, Anan, Tokushima 774-0017, Japan}

\begin{abstract}
Preventing infectious disease like flu from spreading to large communities is one of the most important issues for humans. One effective strategy is voluntary vaccination, however, there is always the temptation for people refusing to be vaccinated because once herd immunity is achieved, infection risk is greatly reduced. In this paper, we study the effect of social impact on the vaccination behavior resulting in preventing infectious disease in networks.
The evolutionary simulation results show that the social impact has both positive and negative effects on the vaccination behavior.
Especially, in heterogeneous networks, if the vaccination cost is low the behavior is more promoted than the case without social impact.
In contrast, if the cost is high, the behavior is reduced compared to the case without social impact.
Moreover, the vaccination behavior is effective in heterogeneous networks more than in homogeneous networks.
This implies that the social impact puts people at risk in homogeneous networks.
We also evaluate the results from the social cost related to the vaccination policy.
\end{abstract}


\end{frontmatter}


\section{Introduction}
\label{intro}
From the past to the present, infectious diseases have threatened humans all over the world \cite{SARS2004, Fouchier_etal2003, A_INFUL2005, Small_etal2007, S_INFUL2009}.
One of the effective ways to prevent diseases is voluntary (preemptive) vaccination.
However, the behavior has a social dilemma structure.
Basically, people can prevent diseases by getting vaccinated but hesitate to do so because they will bear the cost.
Nevertheless, if people do not get vaccinated, they may be infected, which is worse than paying the vaccination cost.
Here, once ``herd immunity'' \cite{Fine_etal2011} is achieved in communities, not getting vaccinated is more attractive to people because the risk to be infected is greatly reduced in such a situation.
Thus, people want other people to get vaccinated but do not want to be vaccinated themselves.
People want a free ride on other people's cooperative behavior.
Because of this dilemma, it is difficult to spread the voluntary vaccination.

Many studies have been done to investigate situations which promote voluntary vaccination.
First, game theoretic models with epidemic dynamics have been used to explain the individual vaccination in well-mixed populations \cite{Bauch_etal2003, BauchEarn2004, Vardavas_etal2007, Breban_etal2007}, random networks \cite{PerisicBauch2009a, PerisicBauch2009b}, or scale-free networks \cite{Zhang_etal2010}.
The game theoretic analysis commonly assumes that the strategy for vaccination is static.
However, in reality, it can be changed over time as the epidemic spreads.
Then, evolutionary game theoretic models combined with epidemiological dynamics classified as ``evolutionary vaccination games'' have been developed \cite{Bauch2005, BhattacharyyaBauch2010, BauchBhattacharyya2012, d'Onofrio_etal2012, Wells_etal2013}.
Since real human interactions can be described by networked populations, some studies focused on evolutionary vaccination games using heterogeneous networks \cite{Fu_etal2011, Cardillo_etal2013, NdeffoMbah2012, Zhang_etal2013}.
Fu et al.~and Cardillo et al.~revealed that vaccination is effective despite the high cost in heterogeneous networks compared to homogeneous networks, because once hub nodes are vaccinated, the behavior immediately spreads to the whole population \cite{Fu_etal2011, Cardillo_etal2013}.
Cardillo et al.~, however, have shown that if the vaccine is imperfect, homogeneous networks conversely outperform heterogeneous networks \cite{Cardillo_etal2013}.
Recently, Han and Sun considered a dynamic model of the evolutionary vaccination game in which the relationship among individuals is dynamically changed over time \cite{HanSun2016}.
Another study focuses on multiple networks in the evolutionary vaccination game \cite{Fukuda_etal2015}.

In those networked models, when imitation of the strategy is conducted, only the payoff information of one particular neighbor is obvious.
If people use broader information, what happens to the dynamics?
Based on this motivation, several studies used various types of information such as, social status information \cite{Fukuda_etal2014}, memory \cite{Zhang_etal2012}, memory and conformism \cite{HanSun2014}, and other-regarding tendencies \cite{YZhang2013}.
Fukuda et al.~investigated the effect of social status information in which each individual can use the information of the average payoff of its neighbors instead of the payoff of one randomly selected neighbor \cite{Fukuda_etal2014}.
The idea comes from the study of the evolution of cooperation which shows that cooperation is promoted using a social average payoff \cite{Shigaki_etal2012}.
Fukuda et al.~showed that the effectiveness of the social status depends on both network structures and the cost of vaccination.
In a similar study, Zhang incorporated the effect of the neighbors' payoffs, called ``other-regarding tendency'', and showed that the tendency influences the vaccination coverage differently depending on the vaccination cost \cite{YZhang2013}.
Moreover, Han and Sun considered the situation that hub players can greatly influence others on networks \cite{HanSun2014}.
They showed that not following such hubs contributes to the large vaccination coverage.

Those studies use the payoff of others as the broader information.
In reality, people sometimes follow their neighbors' decisions irrespective of the payoff.
In other words, people may compare the number of vaccinated neighbors with non-vaccinated neighbors and then imitate the majority of them.
In this paper, we propose a new model in which the effect of such ``social impact'' \cite{Latane1981} is incorporated on networks.
Wu and Zhang have recently developed a similar model called ``peer pressure'' where an individual is influenced by the number of neighbors who have a different strategy \cite{WuZhang2013}.
Our computer simulation results show that, in heterogeneous networks, when the cost of vaccination is low, influential social impact increases the vaccination coverage.
On the other hand, the impact reduces the coverage if the cost is high.
Moreover, since the vaccination coverage of homogeneous networks is lower than that of heterogeneous networks as we will show, following the social impact in homogeneous networks may pose a great threat to society.

\section{Model}
\begin{figure}
\begin{center}
\includegraphics[width=4in]{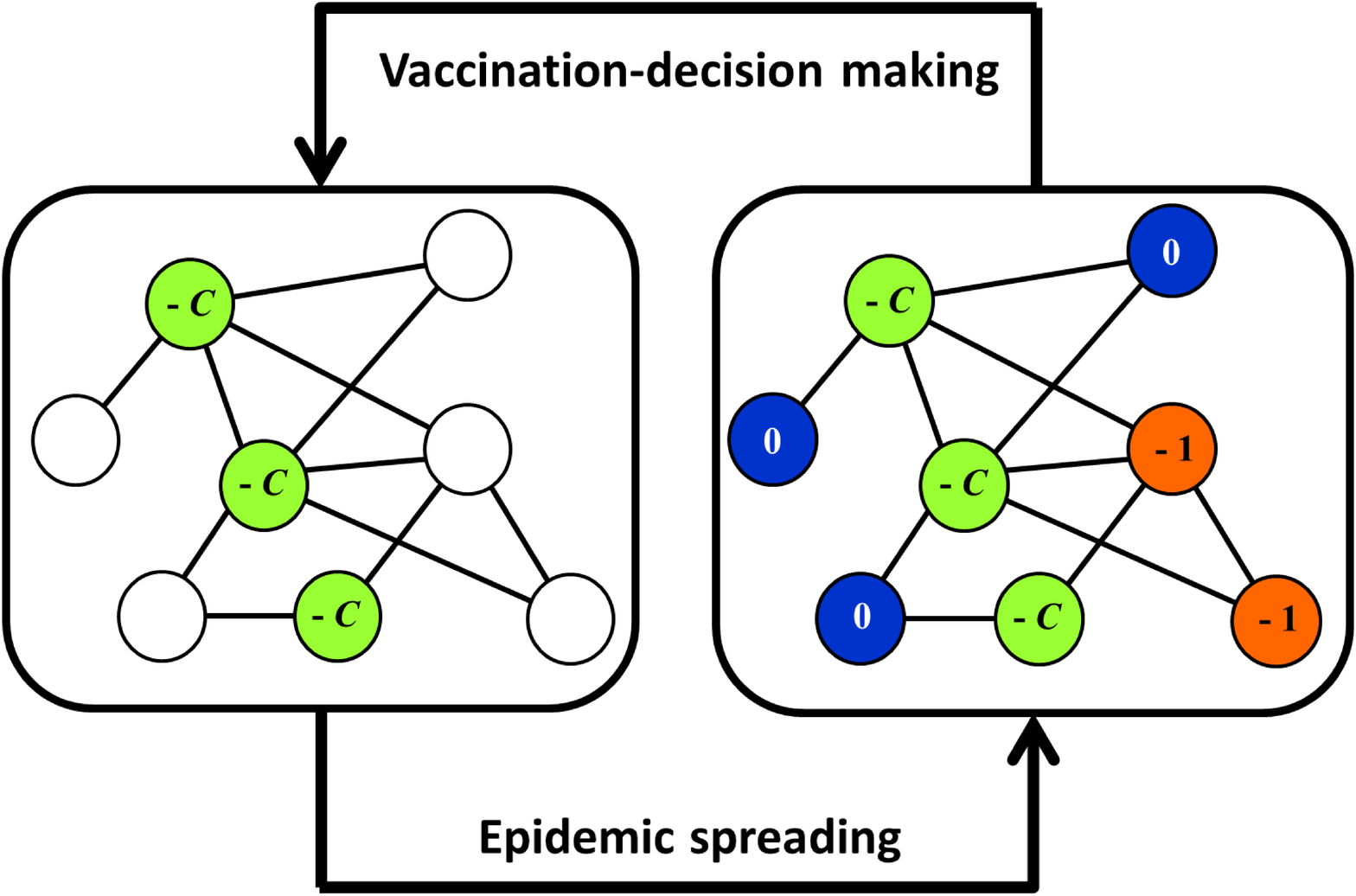}
\caption{Our model consists of two stages: decision-making and epidemic spreading. In the decision-making stage, each individual makes a decision whether to get vaccinated or not. If an individual chooses to get vaccinated, his payoff is $-C$ because of the vaccine cost but results in the perfect immunity against the flu-like disease. After that, the epidemic spreading takes place according to SIR dynamics. If an individual gets infected, his payoff is $-1$. If an individual can avoid the disease, his payoff is 0 because he does not need to pay any cost. The vaccinated individuals never get infected. Since we suppose a seasonal influenza-like disease, it disappears at the end of the season.
At the beginning of the next season, individuals have to make a decision again based on the last year's outcome, and then the next epidemic starts.
These two stages are repeated every year.}
\label{model}
\end{center}
\end{figure}

We developed a model in which an epidemic spreading and individual decision-making take place alternately.
Before the epidemic spreads, individuals make a decision whether to get vaccinated or not based on their vaccination strategy.
The decision is affected by the payoff of others or the social impact, which is controlled by the probability $\alpha$.
Then, epidemic spreading takes place based on the standard susceptible-infectious-recovered (SIR) model.
Here, we assume seasonal influenza as the target disease.
To realize them, we consider two stages: decision-making stage and epidemic spreading stage (Fig.~\ref{model}).
The detail of each dynamic is the following.

\subsection{Decision-making stage}
Each individual makes a decision whether to take vaccination or not.
We assume that the vaccine gives the perfect immunity to vaccinated individuals.
Once an individual decides to get vaccinated, his payoff is $- C$ as the vaccination cost ($0 \le C \le 1$).
The cost may include the monetary cost and the side effects.
When an individual chooses not to get vaccinated, there are two situations.
If the individual is infected, the payoff is $-1$ because getting infected is the worst case.
In contrast, if the individual is not infected, the payoff is 0 as the ideal case.
We also calculate the payoffs of the neighbors of individual $i$ in the same way.

The probability $P(s_i \gets s_j)$ that the individual $i$ with the strategy $s_i$
imitates $j$'s strategy $s_j$ is given by a pairwise comparison according to the Fermi function \cite{SzaboToke1998, Traulsen_etal2007}.
With probability $1-\alpha$, the focal individual $i$ compares his payoff with one randomly selected neighbor, $j$'s payoff, and then updates his strategy based on Eq.~1 ({\it payoff-based strategy updating}).
On the other hand, with probability $\alpha$, the focal individual $i$ ignores his payoff and neighbors' payoffs and counts the number of his neighbors' strategies, and then updates his strategy based on Eq.~2 ({\it popularity-based strategy updating}).

\begin{equation}
P(s_i \gets s_j)=  \frac{1}{1+\exp[\frac{\pi_i - \pi_j}{\kappa}]}
\end{equation}

\begin{equation}
P(s_i \gets s_j)=  \begin{cases}
    \frac{1}{1+\exp[\frac{N_{non(i)}-N_{vac(i)}}{\kappa}]} &  (N_{vac(i)} \geq N_{non(i)}) \\
    \frac{1}{1+\exp[\frac{N_{vac(i)}-N_{non(i)}}{\kappa}]} & (otherwise)
  \end{cases}
\end{equation}

where $\kappa$ is a measure of noise and controls the strength of selection $(0 < \kappa < \infty)$.
For $\kappa \to 0$, individuals are sensitive to the difference of payoff and popularity.
In this paper, we fixed the value of $\kappa$ to be $\kappa = 0.1$  to be the same as the typical studies \cite{HanSun2014, Fukuda_etal2014}.
$\alpha$ controls the intensity between the payoff factor and the popularity factor $(0 \le \alpha \le 1)$.
$N_{vac(i)}$ ($N_{non(i)}$) is the number of the vaccinated (non-vaccinated) neighbors of $i$.

\subsection{Epidemic spreading stage}
After the individuals have been or not have not been vaccinated by the above decision-process, the epidemic spreading stage takes place.
The dynamics are described by the standard SIR model in which the population is divided into three types: Susceptible ($S$), Infectious ($I$), and Recovered ($R$) individuals.
First, the disease is triggered by randomly selected infectious individuals $I_0$. 
After the disease breaks out, each non-vaccinated individual $i$ gets infected with the probability $1-(1-\beta)^{N_{inf(i)}}$ where ${N_{inf(i)}}$ is the number of infected neighbors and $\beta$ is the transmission rate per day.
$\beta$ is different depending on the network topology as described in the next subsection.
There is no risk of infection for the vaccinated individuals because the vaccination provides perfect immunity.
Once each individual is infected, he recovers from the infection with rate $\gamma$ per day.
We set $\gamma=1/3\, \mathrm{day}^{-1}$.
Since it is difficult to mathematically describe the dynamics of structured populations, we numerically simulate the epidemic dynamics by using the Gillespie algorithm \cite{Gillespie1977, Fu_etal2011}.
The values of $\beta$ and $\gamma$ are determined by assuming the flu-like disease.  

\subsection{Network structure}
The above two stages of dynamics are conducted on networks.
Here, each individual is represented as a node and each edge means that there is an interaction between two individuals.
In this paper, along the same lines as the previous study \cite{Fukuda_etal2014}, we consider three typical networks: a square lattice, a random regular graph (RRG) \cite{Ballobas1985}, and the Barab\'asi-Albert scale-free (BA) network \cite{BarabasiAlbert1999}.
In the supplementary material, we also consider the Erdo\H{o}s-R\'enyi (ER) random graph \cite{ErdosRenyi1959}.
Without vaccination, the same $\beta$ yields the different final epidemic size \cite{Fu_etal2011}.
Thus, with 0.9 as the same final epidemic size for the baseline \cite{Fukuda_etal2014}, we select the different transmission rate for each network: $\beta =0.46\, \mathrm{day}^{-1}\, \mathrm{person}^{-1}$ for the square lattice, $\beta =0.37\, \mathrm{day}^{-1}\, \mathrm{person}^{-1}$ for the RRG, and $\beta =0.55\, \mathrm{day}^{-1}\, \mathrm{person}^{-1}$ for the BA network.

\section{Results and discussions}
We conducted simulation experiments using the individual-based model. 
The parameter setting used for the experiments is as follows unless noted otherwise: population size $N=1600$, average degree $k=4$, recovery rate $\gamma=1/3\, \mathrm{day}^{-1}$, transmission rate $\beta$ is different depending on each network, seeds of disease spreading $I_0 = 5$.
The population size is enough to be evaluated because it is consistent with the results of larger populations \cite{Fukuda_etal2014}.
In the first iteration, equal fractions of the vaccinated and non-vaccinated individuals are randomly located in each network, then the epidemic starts.
From the second iteration, individuals decide their strategy based on the algorithm of the decision-making stage followed by the spread of the epidemic.
The iterations finish at 1000. The last 100 iterations are averaged and used as the equilibrium results in all figures.

\subsection{Effect of social impact on voluntary vaccination}
\begin{figure}
\begin{center}
\includegraphics[width=\textwidth]{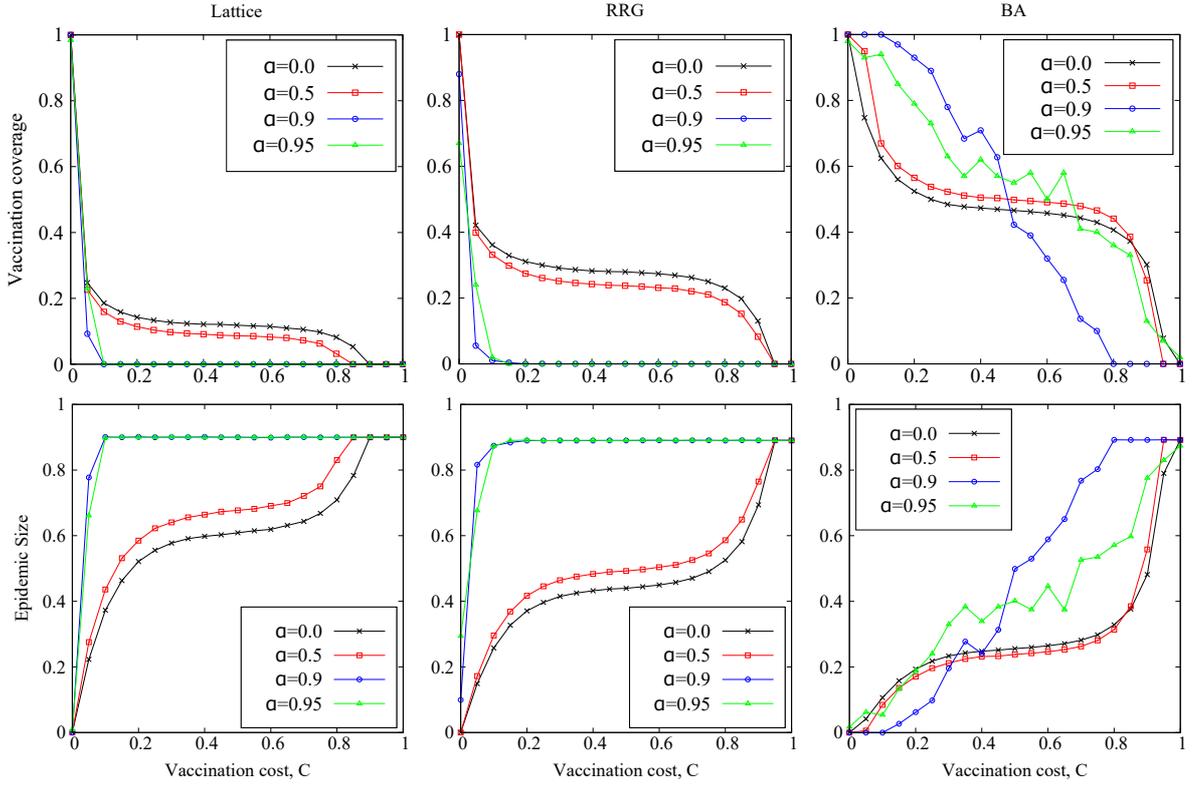}
\caption{Effects of social impact ($\alpha$) to vaccination coverage and final epidemic size as functions of vaccination cost $C$ in different networks.
One hundred independent simulation runs are averaged and the data interval is 0.05.
The results of $\alpha=0$ correspond to the previous studies \cite{Fu_etal2011, Fukuda_etal2014}.
In the square lattice and the RRG, the social impact negatively facilitates the vaccination coverage and the final epidemic size.
In contrast, the BA network both positively and negatively facilitate them.
This is because the majority at the specific vaccination cost is influenced by the social impact.
The entire $\alpha$ dynamics are shown in Fig.~\ref{HM}.
Social impacts $\alpha=0.0, \, 0.50, \, 0.90, \, 0.95$ are shown.
Transmission rate: for each network: $\beta =0.46\, \mathrm{day}^{-1}\, \mathrm{person}^{-1}$ for the square lattice, $\beta =0.37\, \mathrm{day}^{-1}\, \mathrm{person}^{-1}$ for the RRG, and $\beta =0.55\, \mathrm{day}^{-1}\, \mathrm{person}^{-1}$ for the BA network.}
\label{VC}
\end{center}
\end{figure}

We systematically varied values of $\alpha$ to know the effect on the vaccination coverage and the final epidemic size in the different networks.
Figure \ref{VC} shows vaccination coverage and the final epidemic size as the functions of the vaccination cost and the value of $\alpha$.
First, we focus on $\alpha=0$, which means that individuals care only about the payoff difference
with one randomly selected neighbor and do not pay attention to the popularity of the neighbors.
In this case, the BA network most enhanced the vaccination behavior; the RRG was the second; the square lattice was the last.
The results are consistent with the previous ones \cite{Fu_etal2011, Fukuda_etal2014}.
If there is no vaccine, infectious diseases easily spread to the population in the BA network and the RRG compared to the square lattice.
This is because the heterogeneity in the BA network or the randomness in the RRG makes the average path length of the network shorter.
This conversely promotes vaccination coverage since, in this situation, herd immunity is difficult to achieve and then there is less incentive to free-ride by not getting vaccinated.
Therefore, these two networks enhance voluntary vaccination. 

Then, we focus on how the social impact factor $\alpha$ affects voluntary vaccination.
In the square lattice and the RRG populations, the vaccination behavior is greatly suppressed (Fig.~\ref{VC}).
Basically, for these networks, the vaccination coverage is less than 0.5 at $\alpha=0$, which means that getting vaccinated is not popular.
This behavior is negatively facilitated by the social impact and the behavior is promoted further as $\alpha$ becomes larger.
Thus, we can conclude that the social impact is harmful for both networks.
In the ER random graph, we also see the same tendency with these two networks (except for the very low vaccination cost) because getting vaccinated is unpopular within the wide range of costs even without the social impact (see the supplementary material).
We shift to the result of the BA network.
In this case, the social impact has both positive and negative effects on voluntary vaccination.
If the vaccination cost is low ($C_r < 0.4$), the social impact positively facilitates voluntary vaccination because the behavior is already common ($> 0.5$) even without the factor $\alpha=0$.
Conversely, the social impact has the reverse effect  when the vaccination cost is high ($C_r > 0.5$).
The behavior is suppressed similar to the square lattice and the RRG cases.

\begin{figure}
\begin{center}
\includegraphics[width=\textwidth]{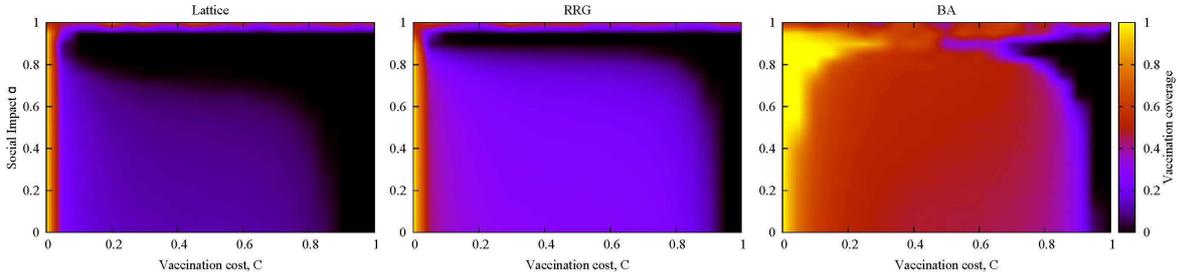}
\caption{The $\alpha$-$C$ phase diagrams of the vaccination coverage in different networks.
Thirty independent simulation runs are averaged and the data interval is 0.05.
If $\alpha$ is the highest ($\alpha=1.0$), individuals do not care about getting infected or the vaccination cost, resulting in random fluctuation of the vaccination coverage. Consequently, the coverage reaches 0.5 on average after 30 independent simulations.}
\label{HM}
\end{center}
\end{figure}

Note that in the BA network, if $\alpha$ is too high, the lower $\alpha$ facilitates a positive or negative effect moreso than the higher value (see the entire $\alpha$ dynamics in Fig.~\ref{HM}).
For example, comparing $\alpha=0.9$ (the lower) with $\alpha=0.95$ (the higher),
the former case easily facilitates both positive and negative effects.
In the case of $\alpha=1.0$, the vaccination cost has no meaning any more.
This leads to random fluctuation and causes almost 0.5 vaccination coverage on average.
Thus, if $\alpha$ is too high, the payoff-based optimization does not work well.
In other words, there is an optimal point of the value of $\alpha$ for the vaccination behavior depending on the balance between the payoff-based and popularity-based imitation behaviors.

\subsection{Social impact affects on total social cost}
\begin{figure}
\begin{center}
\includegraphics[width=\textwidth]{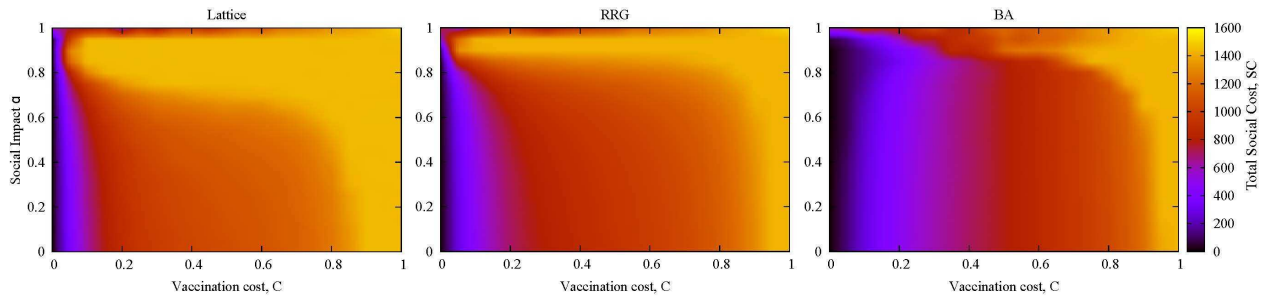}
\caption{The $\alpha$-$C$ phase diagrams of the social cost ($SC$) in different networks.
Thirty independent simulation runs are averaged and the data interval is 0.05, which is interpolated by a smooth function.
The RRG and the square lattice increase $SC$ in every vaccination cost while the BA network reduces $SC$ when the cost is not so high.}
\label{SC}
\end{center}
\end{figure}

It is important to know the total social cost at the population level. The cost includes not only the cost of vaccination but also the cost of infection, which can be defined by \cite{Zhang_etal2013b}

\begin{equation}
SC=N_{R}\times 1.0+N_{V}\times c
\end{equation}

where $N_R$, $N_V$ are the final number of infected (also recovered) and vaccinated individuals, respectively.
For societies, it may be better not to promote vaccination among people even if the number of infected people becomes higher as long as the total cost is kept low.
Figure \ref{SC} shows the phase diagram of the social cost $SC$ depending on the vaccination cost $C$ and the social impact $\alpha$.
Basically, the tendencies are the same with Fig.~\ref{HM}: the social impact gives negative influences (high social cost) at almost all $C$ in the RRG, the square lattice, and the ER network.
Especially, if $\alpha$ is high enough (around $\alpha=0.9$), the negative influence is further promoted.
On the other hand, the social cost has positive (negative) influence when $C$ is low (high) in the BA.
Thus, even at the society level, we can conclude that the social impact works well only in heterogeneous networks to suppress the total social cost.

\subsection{Effect of initial vaccinated individuals}
\begin{figure}
\begin{center}
\includegraphics[width=\textwidth]{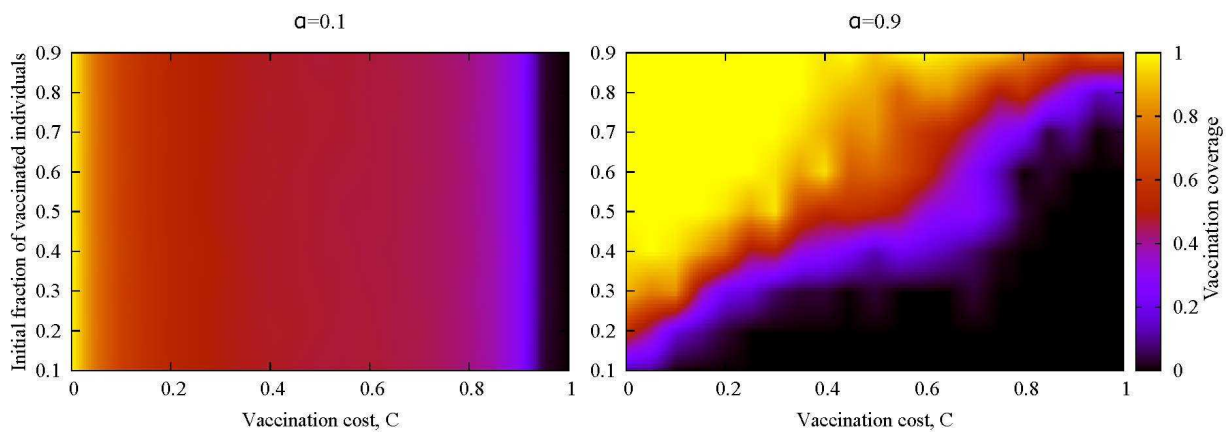}
\caption{Vaccination coverage as functions of the initial fraction of vaccinated individuals and the vaccination cost $C$ in BA networks ($\beta =0.55\, \mathrm{day}^{-1}$) for the different social impacts: (a) $\alpha=0.1$ and (b) $\alpha=0.9$.
Thirty independent simulation runs are averaged and the data interval is 0.1.
When the initial vaccination level is high, the vaccination behavior is promoted in (b) compared to (a) even if the vaccination cost is high because the initial advantage remains unchanged.}
\label{InitialFraction}
\end{center}
\end{figure}

So far, we described that the social impact has positive and negative effects for facilitating voluntary vaccination.
We reasoned that the cases obtained depend on the existing popularity of the population.
To verify this, the initial fraction of the vaccinated individuals is changed in the BA network; most individuals are vaccinated is 0.9 whereas less individuals are vaccinated is 0.1.
Figure \ref{InitialFraction} shows the vaccination coverage in those cases.
When $\alpha=0.1$, the initial fraction of the vaccinated individuals has little effect on the final vaccination coverage.
This is because the individuals mostly care about the payoff difference with others and results in the optimal coverage threshold based on the payoff.
In contrast, the initial fraction heavily affects the final vaccination coverage in the case of high $\alpha$ ($\alpha=0.9$).
In this case, because the initial fraction is larger than 0.5, vaccinated individuals are already the majority.
Then, even if the voluntary vaccination is unfavorable from the viewpoint of the payoff, the behavior is positively facilitated by the social impact due to the initial advantage.
Through this experiment, we found that the social impact certainly has the facilitating effect both in positive and negative ways.

\section{Conclusions}
In this paper, we investigated the effect of social impact on voluntary vaccination and the final epidemic size in a network population model and analyzed the result by the total social cost from the level of society.
We found that pursuing popular behavior has both positive and negative effects on the amount of the vaccination.
When the vaccination behavior is the majority in the population, the behavior is promoted further by the social impact.
The reverse result is obtained when the vaccination is minor in the population.
This fact is experimentally confirmed by modifying the initial fraction of the vaccinated individuals.

We also found that the effect of the social impact greatly depends on the network structures.
Basically, the square lattice and the RRG do not promote the vaccination behavior even if the social impact is absent.
Thus, in these networks, the social impact negatively affected the vaccination behavior.
Conversely, in the BA network, the social impact positively facilitates the vaccination behavior as long as the vaccination cost is low.

Similar results have also been obtained in the previous study in which individuals make a decision whether to get vaccinated or not from both payoff-based and popularity-based factors probabilistically \cite{XiaLiu2013}.
The critical difference between our model and their model is the following two points.
First, we considered how the strategy for the vaccination behavior changes over time based on evolutionary games whereas a game-theoretic analysis of cost minimization is used in \cite{XiaLiu2013}.
Secondly, we used various network structures to investigate the spread of the vaccination behavior and found that the results are quite different depending on the structures.
These two situations are not considered in the previous study \cite{XiaLiu2013}.
Another previous study based on evolutionary games has shown the positive and negative effects of the social impact \cite{WuZhang2013}. They call it ``peer pressure.''
In their model, the peer pressure is incorporated as a weighting factor of  payoff-based strategy updating. In this case, it is difficult to separate both updating rules.
They have shown that the positive and negative effects are obtained in any networks (both homogeneous and heterogeneous).
In contrast, we show that the negative effect is dominant in homogeneous networks. This may arise from the fact that we have separately defined the updating rules.

From the viewpoint of a society level, what we showed is that societies should take into account the network structures of the population when they make policies for vaccination.
When societies are composed of homogeneous interactions, people should be careful not to be affected by others because following the majority may put people at risk in such a situation. 
Also, the prevalence of the social impact becomes large when people can easily get information on popular behavior.
Therefore, societies may need to control the information available to the public depending on the situation.






\end{document}